# Black-Box Quantum State Preparation with Inverse Coefficients

Shengbin Wang[1,*], Zhimin Wang[1,*], Runhong He[1], Guolong Cui[1], Shangshang Shi[1], Ruimin Shang[1], Jiayun Li[1], Yanan Li[1], Wendong Li[1], Zhiqiang Wei[2,3,†] and Yongjian Gu[1,†]

[1] College of Physics and Optoelectronic Engineering, Ocean University of China, Qingdao 266100, China

[2] College of Computer Science and Technology, Ocean University of China, Qingdao 266100, China

[3] High Performance Computing Center, Pilot National Laboratory for Marine Science and Technology (Qingdao), Qingdao 266100, China

[*] These authors contribute equally to this work.
[†] Correspondence author, e-mail: guyj@ouc.edu.cn; weizhiqiang@ouc.edu.cn

## ABSTRACT

Black-box quantum state preparation is a fundamental building block for many higher-level quantum algorithms, which is applied to transduce the data from computational basis into amplitude. Here we present a new algorithm for performing black-box state preparation with inverse coefficients based on the technique of inequality test. This algorithm can be used as a subroutine to perform the controlled rotation stage of the Harrow-Hassidim-Lloyd (HHL) algorithm and the associated matrix inversion algorithms with exceedingly low cost. Furthermore, we extend this approach to address the general black-box state preparation problem where the transduced coefficient is a general non-linear function. The present algorithm greatly relieves the need to do arithmetic and the error is only resulted from the truncated error of binary string. It is expected that our algorithm will find wide usage both in the NISQ and fault-tolerant quantum algorithms.

## I. INTRODUCTION

Black-box quantum state preparation, first developed by Grover [1], is a widely used quantum computational primitive. It can be taken as input to higher-level quantum algorithms or invoked as intermediate subroutine to transform the amplitudes of quantum state [2-9]. Recently, Sanders et al. [10] made an important breakthrough that almost avoids the need to do arithmetic in black-box state preparation and reduces the complexity greatly. Along with the subsequent works inspired by it [11,12], these improvements bring the black-box state preparation and the related quantum algorithms much closer to reality.

Sanders et al. [10] discussed two kinds of state preparation problem, the "linear coefficients" problem and the "root coefficients" problem. The first problem is to transform an amplitude vector $\vec{\alpha} = (\alpha_0, \alpha_1, \ldots \alpha_{d-1})$ into a quantum state close to $\sum_{i=0}^{d-1} \alpha_i |i\rangle$, while the second one is to $\sum_{i=0}^{d-1} \sqrt{\alpha_i} |i\rangle$ (unnormalized). In the present work, we introduce the "inverse coefficients" problem that is to create the state

proportional to $\sum_{i=0}^{d-1} 1/\alpha_i |i\rangle$. The most famous example of "inverse coefficients" problem is the Harrow-Hassidim-Lloyd (HHL) algorithm [2], which is actually to inverse the matrix. In HHL, after the phase estimation stage, the eigenvalues are entangled as $\sum_i \beta_i |u_i\rangle |\lambda_i\rangle$, then the next step is to transform the state into $\sum_i C\beta_i / \lambda_i |u_i\rangle |\lambda_i\rangle$, which is indeed the "inverse coefficients" problem.

The "inverse coefficients" problem of black-box quantum state preparation is typically solved by controlled rotation, which requires quantum computers to calculate the reciprocals and arcsines and results into a major contributor to the complexity [13]. Even by invoking the algorithm for "linear coefficients" problem as subroutine, one still need to do the arithmetic for reciprocal. The reciprocal can be calculated using the Newton-Raphson iteration method [13-15] which brings much complexity.

Here we propose a new method to address the "inverse coefficients" problem. By extending the idea of inequality test introduced in Ref. [10], our method can solve the inverse-coefficients problem with only one multiplication operation. The quantum multiplication operation is much easy than inverse operation, and there exists a vast amount of literature providing various efficient quantum circuits for the multiplication operation [16-18]. Therefore, the present method improves the cost greatly.

The paper is organized as follows. In Section II, the algorithm for solving the "inverse coefficients" problem is presented. In Section III, we generalize the method to form a framework for solving more general nonlinear coefficients of black-box quantum state preparation. Conclusions are given in Section IV.

## II. ALGORITHM FOR INVERSE COEFFICIENTS PROBLEM

In the scenario for black-box state preparation, the amplitudes vector to be loaded are unknown a priori and are supposed to be accessed through an oracle (black-box). More specifically, suppose the amplitude vector is $\vec{\alpha} = (\alpha_0, \alpha_1, ..., \alpha_{d-1})$ with the element $\alpha_i$ being expressed as a binary string, then there exists an oracle $U_o$ giving the following state

$$U_o \left( \frac{1}{\sqrt{d}} \sum_{i=0}^{d-1} |i\rangle |0\rangle^{\otimes n} \right) = \frac{1}{\sqrt{d}} \sum_{i=0}^{d-1} |i\rangle |\alpha_i\rangle . \tag{1}$$

The amplitude elements $\alpha_i$ are encoded coherently in the second register with $n$-bit precision. Note that the dimension of the amplitude vector $d$ may be not a power of 2, so the uniformly superposition state $\sqrt{d^{-1}} \sum_{i=0}^{d-1} |i\rangle$ cannot be created trivially. Since such superposition state is widely used in quantum algorithms, in the appendix we present a simple and efficient quantum circuit for producing such state with at most one round of standard amplitude amplification operation [19]. It is assumed that the oracle $U_o$ can be invoked, as required, any number of times and the complexity is $O(1)$. In practice, oracle $U_o$ would be a QRAM [20] or a quantum circuit.

Starting with Eq. (1), the task of black-box state preparation with inverse coefficients is to produce a target state as

$$|\text{target}\rangle = \frac{1}{\|1/\vec{\alpha}\|_2} \sum_{i=0}^{d-1} \frac{1}{\alpha_i} |i\rangle, \qquad (2)$$

where the $l_2$ norm $\|1/\vec{\alpha}\|_2 = \sqrt{\sum_i (1/\alpha_i)^2}$ is the normalization coefficient. In linear coefficients case, the value of $\alpha_i$ is restricted into the interval of (0, 1), that is, the binary point is located at the left most position of the binary string (0.$\alpha_i$). However, in inverse coefficients case the target amplitude $1/\alpha_i$ should be scaled into (0, 1), so the invoked $\alpha_i$ is assumed to be a positive integer here. That is to say, the binary point is located at the right most position of the binary string ($\alpha_i$.0).

The basic idea of our method comes from the simple multiplication relation between the input $x$ and its inverse $1/x$, i.e. $x \cdot (1/x) = 1$. If the inverse $1/x$ is replaced by a smaller number $y < 1/x$, then the multiplication satisfies $x \cdot y < 1$; If the inverse $1/x$ is replaced by a larger number $y$, the multiplication satisfies $x \cdot y > 1$. Naturally, the relation $x \cdot y < 1$ or $x \cdot y > 1$ is equivalent to the inequality test in linear coefficients case [10]. So our method is to perform the inequality test $\alpha_i \cdot j < 1$ where $j$ is the computational basis in a uniformly superposition state. Count the number of $j$ that satisfies the inequality and it is approximated to the value $1/\alpha_i$.

As it is in the HHL algorithm, a normalization constant $C$ that is no more than the minimum of all elements, i.e. $C \leq \alpha_{\min}$, can be induced to increase the amplitude in practice. Then the inverse turns to $C/\alpha_i$ and the inequality test turns to $\alpha_i \cdot j < C$. Same as $\alpha_i$, $C$ is chosen to be a positive integer, then $C/\alpha_i$ is a fraction belonging to (0, 1]. So the uniformly distributed $j$ is a positive real number. The circuit representation of our method is shown in Fig. 1, and the details of the state evolution is as follows.

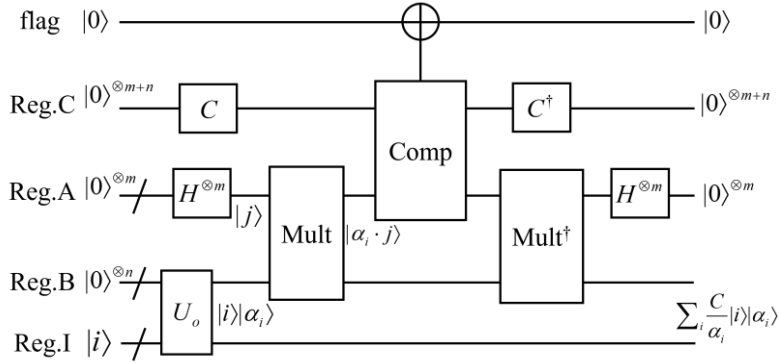

Fig. 1 The overall circuit for transducing the inverse coefficients into amplitude. The module Mult performs the multiplication operation, and Comp is used to test the inequality. The initial data $|\alpha_i\rangle$ is invoked from the oracle $U_o$ based on Eq. (1).

The algorithm consists of the following five steps:
(1) First is to initialize the registers. Assume the data $\alpha_i$ in Reg. B invoked from $U_o$

is truncated in *n*-bit precision; prepare a uniformly superposition state on Reg. A using *m* parallel Hadamard gates; encode the constant *C* in Reg. C with *m+n* qubits. Then we have

$$|\psi_{\text{initi}}\rangle = \left(\frac{1}{\sqrt{d}}\sum_{i=0}^{d-1}|i\rangle_I|\alpha_i\rangle_B\right)\left(\frac{1}{\sqrt{2^m}}\sum_{j=0}^{2^m-1}|j\rangle_A\right)|C\rangle_C. \quad (3)$$

(2) Perform multiplication operation between Reg. B and Reg. A to obtain

$$|\psi_{\text{multi}}\rangle = \frac{1}{\sqrt{d}}\sum_{i=0}^{d-1}|i\rangle_I|\alpha_i\rangle_B\left(\frac{1}{\sqrt{2^m}}\sum_{j=0}^{2^m-1}|\alpha_i\cdot j\rangle_A\right)|C\rangle_C. \quad (4)$$

(3) Compare the value of the bit-string of Reg. A with that of Reg. C, and record the result into the flag qubit. As mentioned above, the position of the binary point should be taken into consideration. Specifically, if $\alpha_i\cdot j < C$, namely $j < C/\alpha_i$, then the indication qubit is $|0\rangle_{\text{flag}}$; if $\alpha_i\cdot j \geq C$, namely $j \geq C/\alpha_i$, then the indication is $|1\rangle_{\text{flag}}$. After this step the state evolves into

$$|\psi_{\text{comp}}\rangle = \frac{1}{\sqrt{d}}\sum_{i=0}^{d-1}|i\rangle_I|\alpha_i\rangle_B\frac{1}{\sqrt{2^m}}\left(\sum_{j=0}^{2^mC/\alpha_i-1}|\alpha_i\cdot j\rangle_A|0\rangle_{\text{flag}} + \sum_{j=2^mC/\alpha_i}^{2^m-1}|\alpha_i\cdot j\rangle_A|1\rangle_{\text{flag}}\right)|C\rangle_C. \quad (5)$$

(4) Undo the multiplication operation on Reg. B and Reg. A, and reset Reg. C. Apply another *m* parallel Hadamard gates on Reg. A to extract $|0\rangle_A^{\otimes m}$ basis $2^m C/\alpha_i$ times indicated by $|0\rangle_{\text{flag}}$, and then we have

$$|\psi_{\text{undo}}\rangle = \frac{1}{\sqrt{d}}\sum_{i=0}^{d-1}|i\rangle_I|\alpha_i\rangle_B\left(\frac{C}{\alpha_i}|0\rangle_A^{\otimes m}|0\rangle_{\text{flag}} + |\omega\rangle_{A\otimes\text{flag}}\right), \quad (6)$$

where $|\omega\rangle$ is an unnormalized state orthogonal to $|0\rangle_{A\otimes\text{flag}}$.

(5) Carry out measurement on Reg. A and flag qubit, if the result is $|0\rangle_{A\otimes\text{flag}}$, then the state of Regs. I and B is

$$|\psi_{\text{succ}}\rangle = \frac{1}{\|1/\vec{\alpha}\|_2}\sum_{i=0}^{d-1}\frac{1}{\alpha_i}|i\rangle_I|\alpha_i\rangle_B. \quad (7)$$

Finally reset Reg. B using the inverse oracle $U_o^{-1}$, then we obtain the target state $\frac{1}{\|1/\vec{\alpha}\|_2}\sum_{i=0}^{d-1}\frac{1}{\alpha_i}|i\rangle_I$ in Reg. I. Note that, in practice, before measurement the amplitude amplification [19,21] can be performed to boost the success probability to get the target state. The number of rounds of amplitude amplification is $O(\sqrt{d}/\|C/\vec{\alpha}\|_2)$.

The above approach can be adapted to perform the division operation easily. In this case, the inputs are positive real values $\alpha_i$ and $\beta_i$, and the target is to transduce the division $\beta_i/\alpha_i$ into the amplitude. We assume $\beta_i \leq \alpha_i$, so the division belongs to the interval $(0,1]$. Other values of $\beta_i/\alpha_i$ should be first scaled into $(0,1)$. In order to perform the division $\beta_i/\alpha_i$, one only need to initialize the Reg. C in Fig. 1 by $\beta_i$, and the result is finally encoded as $|\psi_{\text{div}}\rangle = \frac{1}{\|\vec{\beta/\alpha}\|_2} \sum_{i=0}^{d-1} \frac{\beta_i}{\alpha_i} |i\rangle_I |\alpha_i\rangle_B$.

Now we analyze the complexity and error of our approach. We compare the complexity in the terms of the number of multiplication operations. Typically, the reciprocal is calculated by the Newton-Raphson iteration method [13]. The number of iterations is $\log_2\log_2(1/\varepsilon)$ with $\varepsilon$ being the error of the result. In each iteration step of Newton-Raphson method, there contains two multiplication operations forward, and the un-computation of the entire forward multiplication operations needs to be done to eliminate the garbage qubits after $1/\alpha_i$ is transduced into amplitude. That is the total number of multiplications to reach the result precision $1-\varepsilon$ is about $4\log_2\log_2(1/\varepsilon)$ based on Newton-Raphson method while our method, as shown in Fig. 1, needs only 2 multiplications and can reach arbitrary precision. The precision of our method is simply bounded by the precision of $j$ in Reg. A, i.e. $2^{-m}$, when the data $\alpha_i$ is assumed to be precise. If $\alpha_i$ is regarded as a $n$-bit truncated input, $m$ can also be set to guarantee a much smaller error induced by the truncation of $\alpha_i$. Additionally, the Newton-Raphson method needs to prepare an appropriate initial state and $\log_2\log_2(1/\varepsilon)$ more times qubits.

A significant difference between the Newton-Raphson iteration method and the present one should be pointed out further: the effect of the un-computation process. The state of $|1/\alpha_i\rangle$ is approximated by the forward multiplication operations using Newton-Raphson method, and un-computed back to $|0\rangle$ at the end. This un-computation operation is just used to eliminate the intermediate garbage qubits. In contrast, the present method picks out the first number of $2^m C/\alpha_i$ basis indicated by $|0\rangle_{\text{flag}}$ in the uniformly superposition state in Reg. A by the modules Mult. and Comp. Up to now, the amplitude of $|0\rangle_{\text{flag}}$ is $\sqrt{1/\alpha_i}$, rather than $1/\alpha_i$, with an intermediate garbage state $\sum_{j=0}^{2^m C/\alpha_i - 1} |\alpha_i \cdot j\rangle_A$, see Eq. (5). Until Mult$^+$ and another $m$ parallel Hadamard operations (the un-computation) are performed in Reg. A, see Eq. (6), the residual proportional amplitude of $\sqrt{1/\alpha_i}$ indicated by $|0\rangle_A^{\otimes m}$ is invoked from the intermediate state. Finally, the target $1/\alpha_i$ is transduced into amplitude. So the un-computation is used to not only eliminate the intermediate garbage qubits but also calculate $1/\alpha_i$.

### III. EXTENSION OF THE ALGORITHM

We can extend our approach to be a framework for solving general problems of black-box quantum state preparation. In this case, starting with Eq. (1), one would aim to create such a state

$$|\text{target}\rangle = \frac{1}{\|f(\vec{\alpha})\|_2} \sum_{i=0}^{d-1} f(\alpha_i)|i\rangle, \qquad (8)$$

where $f(x)$ is a non-linear function and $\|f(\vec{\alpha})\|_2 = \sqrt{\sum_{i=0}^{d-1} f^2(\alpha_i)}$ is the normalization coefficient.

Typically, this general problem would be done through two steps, first performing the arithmetic $|\alpha_i\rangle|0\rangle \mapsto |\alpha_i\rangle|f(\alpha_i)\rangle$, and then transducing the amplitude by the approach of black-box state preparation with linear coefficients. Obviously, here the forward map from $\alpha_i$ to $f(\alpha_i)$ is performed to get the function values. In contrast, in the above algorithm for the inverse coefficients problem, the reciprocal is calculated utilizing the backward map. That is, in Fig. 1, first all possible approximations of reciprocal $1/\alpha_i$ are prepared in the computational basis of the uniformly superposition state on Reg. A, and then the target value is picked out by multiplying them with $\alpha_i$ and performing inequality test. It implies that combining the inverse map and the inequality test would result into a better way of transducing the $f(x)$ values. Specifically, without loss of generality, supposing that $f(x)$ is expressed as a composite function, namely $y = f(x) = h(g(x))$, first we prepare all possible values of $h^{-1}(y)$ and $g(x)$ on two registers, respectively, and then perform inequality test between the two registers to pick out the function value $y$.

Our extension algorithm is shown schematically in Fig. 2. The state evolution is similar as that of Fig. 1. The unitaries $U_{h^{-1}}$ and $U_g$ are used to perform the $h^{-1}(y)$ and $g(x)$ functions, respectively. After this step, the state is

$$|\psi_U\rangle = \frac{1}{\sqrt{d}} \sum_{i=0}^{d-1} |i\rangle_\text{I} |\alpha_i\rangle_\text{B} |g(\alpha_i)\rangle \left( \frac{1}{\sqrt{2^m}} \sum_{j=0}^{2^m-1} |j\rangle_\text{A} |h^{-1}(j)\rangle \right). \qquad (9)$$

Then after the comparison operation, the state evolves into

$$|\psi_\text{comp}\rangle = \frac{1}{\sqrt{d}} \sum_{i=0}^{d-1} |i\rangle_\text{I} |\alpha_i\rangle_\text{B} |g(\alpha_i)\rangle \frac{1}{\sqrt{2^m}} \left( \begin{array}{l} \sum_{h^{-1}(j)<g(\alpha_i)} |j\rangle_\text{A} |h^{-1}(j)\rangle |0\rangle_\text{flag} \\ + \sum_{h^{-1}(j)\geq g(\alpha_i)} |j\rangle_\text{A} |h^{-1}(j)\rangle |1\rangle_\text{flag} \end{array} \right). \qquad (10)$$

Finally, applying uncomputation and parallel $H$ gates, the target state shown in Eq. (8) is obtained.

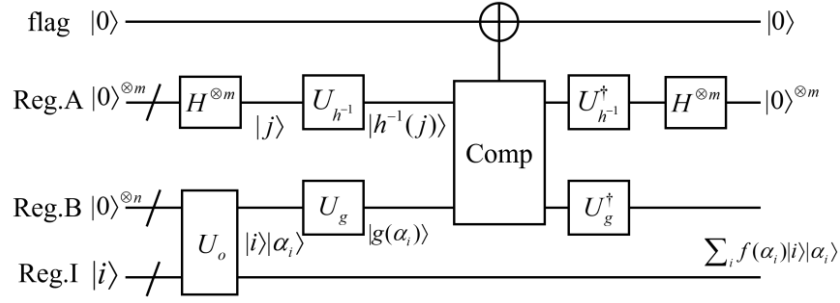

Fig. 2 The overall circuit for transducing the general coefficients, i.e. $f(α_i)$, into amplitude.

Now we take the function $f(\alpha_i) = 1/\sqrt{1+\alpha_i}$ as an example to further explain our algorithm. In this case, the forward map is $g(\alpha_i) = 1+\alpha_i$ and the backward map is $h^{-1}(j) = j^2$, and they are evaluated in parallel on Reg. A and B, respectively. Then perform multiplication between them and use inequality test of $g(\alpha_i) \cdot h^{-1}(j) < 1$ to pick out the corresponding basis $j$ as it is done in the case of inverse coefficients. Therefore, in our approach the sequential square root and inverse operations to evaluate $f(α_i)$ are replaced by parallelized square and multiplication operations. The complexity is reduced greatly. The precision of the resulted amplitude, i.e. $f(α_i)$, is also bounded by the precision of $j$ in Reg. A, i.e. $2^{-m}$, when $α_i$ is a precise data.

We expect that more concrete functions will be found that can be transduced into amplitude using our approach. More exploration of practical usage of our algorithm is also necessary. We leave these for the future work.

## IV. CONCLUSIONS

In this work, we present an algorithm for black-box quantum state preparation with inverse coefficients based on inequality test method. The reciprocal information is first encoded in a uniformly superposition state, then through just two multiplication operations and one inequality test the target reciprocal can be transduced into amplitude. Furthermore, we extend this method for general problem of black-box state preparation with a general non-linear function $f(x)$ being the coefficients. By expressing $f(x)$ as $y = f(x) = h(g(x))$, the forward map $g(x)$ and backward map $h^{-1}(y)$ can be evaluated on two registers in parallel, then the inequality test is applied to pick out the target value $f(x)$. The error is mainly the truncated error because of the finite qubits of registers. Our algorithm relieves the need of arithmetic greatly, although cannot avoid it, and makes significant reductions to the complexity of performing general black-box state preparation.

In quantum computing, the information can be encoded in basis, phase and amplitude, and can be transduced to each other among them [13-15,19,22-24]. Here we discuss the conversion from basis to amplitude. We expect that combining the present algorithm with other conversion approaches will result into a more versatile tool for quantum information processing.

## ACKNOWLEDGEMENTS

The present work is supported by the National Natural Science Foundation of China (Grant No. 12005212 and 61575180) and the Pilot National Laboratory for Marine Science and Technology (Qingdao).

## APPENDIX

In the appendix, we present the approach to prepare the uniformly superposition state $|\psi_{\text{unif}}\rangle = \sqrt{d^{-1}} \sum_{i=0}^{d-1} |i\rangle$. When $d$ is a power of 2, i.e. $d=2^l$, $l$ parallel Hadamard gates are enough to produce the state $|\psi_{\text{unif}}\rangle$. For the general case that $2^{l-1}<d<2^l$, the state can be created based on the inequality test and amplitude amplification as schematically shown in Fig. 3.

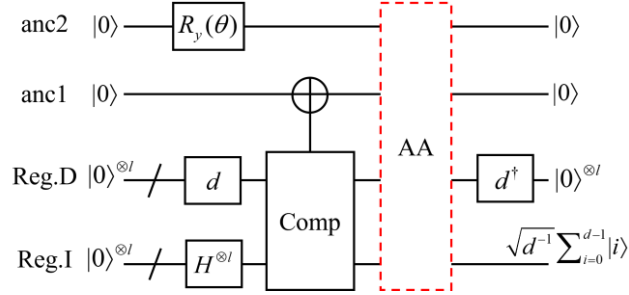

Fig. 3 The overall circuit to prepare the general uniformly superposition state with the number of computational basis being not a power of 2. AA represents the standard amplitude amplification.

The procedure of this algorithm goes as follows:

(1) Apply $l = \lceil \log d \rceil$ Hadamard gates on Reg. I, and prepare value $d$ on Reg. D. Then we have the state

$$|\psi_{\text{I}}\rangle = \frac{1}{\sqrt{2^l}} \sum_{i=0}^{2^l-1} |i\rangle_{\text{I}} |d\rangle_{\text{D}} . \tag{11}$$

(2) Perform the comparison operation between Regs. D and I, and mark the relation on ancillary qubit anc1. Then the state evolves into

$$|\psi_{\text{comp}}\rangle = \frac{1}{\sqrt{2^l}} |d\rangle_{\text{D}} \left( \sum_{i=0}^{d-1} |i\rangle_{\text{I}} |0\rangle_{\text{anc1}} + \sum_{i=d}^{2^l-1} |i\rangle_{\text{I}} |1\rangle_{\text{anc1}} \right). \tag{10}$$

(3) Append another qubit anc2 and perform $R_y(\theta)$ rotation with $\theta$ being $2\arccos\sqrt{2^{l-2}/d}$, then we obtain

$$|\psi_{R_y}\rangle = \frac{1}{\sqrt{2^l}}|d\rangle\left(\sum_{i=0}^{d-1}|i\rangle|0\rangle_{anc1} + \sum_{i=d}^{2^l-1}|i\rangle|1\rangle_{anc1}\right)\left(\sqrt{\frac{2^{l-2}}{d}}|0\rangle_{anc2} + \sqrt{1-\frac{2^{l-2}}{d}}|1\rangle_{anc2}\right). \quad (11)$$

The $R_y(\theta)$ rotation is used to adjust the magnitude of the amplitude to set the stage for the following amplitude amplification. The target state is indicated by $|0\rangle_{anc1}|0\rangle_{anc2}$, that is, Eq. (11) can be rewritten as

$$|\psi_{R_y}\rangle = \frac{1}{2}\left(\sum_{i=0}^{d-1}\frac{1}{\sqrt{d}}|i\rangle\right)|d\rangle|0\rangle_{anc1}|0\rangle_{anc2} + |\omega\rangle, \quad (12)$$

where $|\omega\rangle$ contains the states orthogonal to $|0\rangle_{anc1}|0\rangle_{anc2}$. Now the probability of the target state is 1/4, hence just one round of standard amplitude amplification is enough to precisely increase the success probability to be one [19].

Note that in practice the comparator can be designed quite simple because the binary $|d\rangle$ is known beforehand. The qubits and non-Clifford Toffoli gates can be reduced in different schemes [18,25,26]. This algorithm has linear qubit and gate complexity with small constant factor and needs no measurement operations which promise its widely unrestricted applications.

According to the normalization property, our algorithm can produce a perfect $|\psi_{unif}\rangle$. However the amplitude of $|0\rangle_{anc1}|0\rangle_{anc2}$ cannot be amplified to be perfect 1 because the value of angle $\theta$ cannot be obtained precisely in most cases. In the following we analyze the impact of the error induced by the angle $\theta$ to the success probability of obtaining the target $|\psi_{unif}\rangle$.

Since $2^{l-1} < d < 2^l$ and $\theta = 2\arccos\sqrt{2^{l-2}/d}$ for $R_y(\theta)$ rotation, the value of $\cos(\theta/2)$ belongs to $(1/2, 1/\sqrt{2})$ and the absolute value of its derivative belongs to $(\sqrt{2}/2, \sqrt{3}/2)$. Suppose the error of rotation angle $\theta$ is $\varepsilon_0$, then we have the following inequality,

$$\frac{\varepsilon_0}{4} < \frac{\sqrt{2}}{2}\frac{\varepsilon_0}{2} \leq \cos\frac{\theta}{2} - \cos\left(\frac{\theta - \varepsilon_0}{2}\right) \leq \frac{\sqrt{3}}{2}\frac{\varepsilon_0}{2} < \frac{\varepsilon_0}{2}. \quad (13)$$

then the error induced into the factor $1/2 = \sin(\pi/6)$ is bounded as

$$\frac{\varepsilon_0}{8} < \sqrt{\frac{d}{2^l}}\frac{\varepsilon_0}{4} < \sqrt{\frac{d}{2^l}}\left[\cos\frac{\theta}{2} - \cos\left(\frac{\theta - \varepsilon_0}{2}\right)\right] < \sqrt{\frac{d}{2^l}}\frac{\varepsilon_0}{2} < \frac{\varepsilon_0}{2}, \quad (14)$$

the angle $\theta_{1/2}$ corresponding to 1/2 lays on the small interval of $\pi/6$ which has a derivative approaching $2/\sqrt{3}$ in practice. That is to say, the angle $\theta_{1/2}$ involves an error belonging to ($\varepsilon_0/8$, $\varepsilon_0$).

Finally, by one round of standard amplitude amplification, the final angle belongs to the interval ($\pi/2-3\varepsilon_0$, $\pi/2-3\varepsilon_0/8$), leading to the final value ranging from $\cos(3\varepsilon_0)$ to $\cos(3\varepsilon_0/8)$. On the interval ($\pi/2-3\varepsilon_0$, $\pi/2+3\varepsilon_0$), the derivative of the sine function approaches 0 which means the impact of $\varepsilon_0$ in $\sin(\pi/2)$ is much smaller, i.e. $|\sin x_1 - \sin x_2| \ll |x_1 - x_2|$. We employ the second order truncated Taylor series of cosine function $\cos x > 1-x^2/2$ to upper bound the error $\varepsilon_u$ in the final amplitude of $|0\rangle_{anc1}|0\rangle_{anc2}$ as

$$\varepsilon_u \leq 1 - \cos(3\varepsilon_0) < \frac{(3\varepsilon_0)^2}{2} < 2^3 \varepsilon_0^2. \tag{15}$$

So the amplitude of $|0\rangle_{anc1}|0\rangle_{anc2}$ is boosted to at least $1 - 2^3\varepsilon_0^2$. That is the probability of obtaining the perfect initial state $|\psi_{unif}\rangle = \sqrt{d^{-1}}\sum_{i=0}^{d-1}|i\rangle$ is more than $1 - 2^4\varepsilon_0^2$.